# Design and Implementation of Database Independent Auto Sequence Numbers

Kisor Ray

*Techno India Agartala, A Degree Engineering College, Affiliated to Tripura University*
*Maheshkhola,Aagartala,Tripura,India*

***Abstract—*** *Developers across the world use autonumber or auto sequences field of the backend databases for developing both the desktop and web based data centric applications which is easier to use at the development and deployment purpose but can create a lot of problems under varied situations. This paper examines how a database independent autonumber could be developed and reused solving all the problems as well as providing the same degree of easy to use features of autonumber offered by modern Relational Database Systems..*

## I. INTRODUCTION

Developers like to use readily available autonumber [1] /auto sequence feature offered by the modern Relational Database Systems(RDBMS). This technique could be very useful while development of a prototype of an application but may not be a very useful one for the overall operational purposes. The use of this feature may become a costlier option in terms of effort specially while migrating[2] from one RDBMS to other. This paper will show how a generic auto sequence/autonumber module can be created once and used repeatedly for all types of relational database tables irrespective of any specific one

.

## II. BACKGROUND

When we planned to develop a Management Information System (MIS) for our budding institute, we initially used MS Access to develop a couple of modules covering the students, staff members and finance management. After the first phase of development, we decided to use open source PostgreySQL 9.4  running under Linux (Ubuntu 12.04 LTS) as our backend database keeping in mind the need of the bigger repository at free of cost as well as licensing hassles . While migrating our database initially developed on MS Access 2010, we found that all the tables which had autonumber associated with needed to be manually modified under the  PostgreySQL database. We also found that the bound  forms with autonumber are relatively difficult to manage specially while adding a new record. So, we decided to design and implement a generic autonumber  module which could be easily called and reused as and when required discarding the database driven default autonumber/auto sequence feature used earlier.

## III. DESIGN CONCEPT

The autonumber module  opens the database table in the backend database  and looks for any pre-existing record. Should  records exist, move the  record counter to the last record and count the total number of records for that particular table .We have considered two variables here to compare , one is the  maximum position index (MaxPI) (which also could be termed as max sequence index) and other is total record count (TRC). MaxPI  is the current  serial no (SL) or identification no (ID)  or any similar number which we are trying to test and compare for the purpose of obtaining the   new number through increment. If  total record count (TRC) is greater or equal to the MaxPI then the new autonumber to be used is TRC+1 , however if TRC is less than the MaxPI then the new autonumber to be used is MaxPI+1.

## IV. CODES

```
'******** Define the Global Variables ********

Option Compare Database
' *** Define Public global variables  outside the  sub ***
'Public UserGID As Integer
Public GMyAutoNum As Integer
Public GMyTab As String
Public GMyTID As String

Public Sub GetAutoNum()

' **** Generate Autonumber/Sequence *******

'***** Define the Local Variables ***

Dim RC As Variant
Dim MaxPI As Variant
Dim TRC As Variant
Dim myAutoVar2 As Variant
Dim myAutoVar3 As Variant

'***** Define the Database *********

Dim AutNumDb As DAO.Database
Dim AutNumRst As DAO.Recordset

'******Get the database table name *****
myAutoVar2 = GMyTID
```

```
' **** Set column name which use autonumber *****

 Set AutNumDb = CurrentDb
 Set AutNumRst = AutNumDb.OpenRecordset(" Select *
from " & GMyTab & " ;", dbOpenDynaset, dbSeeChanges)

  If    AutNumRst.EOF <> True Then
          AutNumRst.MoveLast
          MaxPI = AutNumRst("" & myAutoVar2 & "")
          TRC = AutNumRst.RecordCount

' ***** Compare and Set Receipt Number *****

          If TRC >= MaxPI Then
              myAutoVar3 = TRC + 1
             Else
               If TRC < MaxPI Then

              myAutoVar3 = MaxPI + 1
            End If
         End If

          AutNumRst.Close
          AutNumDb.Close
            Else
      MsgBox " No existing ID or Serial number found!"
             myAutoVar3 = TRC + 1

       End If

' **** Assign generated number to public variable *****

         GMyAutoNum = myAutoVar3

'***** Close database *********

         AutNumRst.Close
         AutNumDb.Close
         Set AutNumRst = Nothing
         Set AutNumDb = Nothing

End Sub
```

## V. IMPLEMENTATION

We have defined a couple of global variables like :
Public GMyAutoNum As Integer
Public GMyTab As String
Public GMyTID As String

All the global variables here starts with 'G'. The variable GMyAutoNum is the autonumber /auto sequence gets generated from the module each time called by the application, GMyTab is the table name which could be any table name from the backend database which uses the auto number/auto sequence , GMyTID is the autonumber/auto sequence filed name of a table used by the application.

Suppose, we have a table Tab_T_Fees which is a transaction table used to collect various fees where one of the objective is to generate sequential unique 'Receipt Number' . The field 'Rcpt_SL' is a numeric field and also serves as a primary key. All we need to do is that in the frontend form, we put the following code :

```
' **** set the table name ****

GMyTab = "Tab_T_Fees"

'**** set the Column name ****
GMyTID = "Rcpt_SL"

'**** Call the Module/Procedure ****
GetAutoNum   ' *** this is the name of the module ***
Me.TxtRcptNum = GMyAutoNum
```
Where, TxtRcptNum is the field name of the form where autonumber / auto sequence is used.

| Field Name | Data Type |
|---|---|
| Roll_Number | Number |
| Receipt_Date | Date/Time |
| Fee | Number |
| Student_Full_Name | Text |
| Fee_Purpose | Text |
| Remark | Text |
| Rcpt_SL | Number |
| Branch | Text |
| Session | Text |
| Duplicate | Text |
| Fee_Purpose_Type | Number |

Fig 1: Transaction table Tab_T_Fees

Fig 2 : Fee Collection and Receipt Generation Form , Receipt No. RCPT_SL gets generated when receipt is printed.

Initially there is no fee payment, so no record exists. Hence, Total Reord Count (TRC) =0 ,RCPT_SL=0. Say, after five transactions, there are five records and RCPT_SL will be 1,2,3,4,5 etc. repectively. For the testing purpose, now let us delete the transaction no. 4 ( that is RCPT_SL 4) . So, we have the following transactions now with serial numbers assigned to the field RCPT_SL like :

Transaction #1 , RCPT_SL =1
Transaction #2 , RCPT_SL =2
Transaction #3 , RCPT_SL =3
Transaction #4 , RCPT_SL =5

Now, let us make a new transaction. In this situation, the MaxPI=5 and TRC=4 , since TRC is less than the MaxPI, the new serial number ( generated through the autonumber/auto sequence module) for the new transaction will be MaxPI+1 = 6 . That means, any deleted number within the sequences could be taken care of because it cannot be reproduced and or manipulated. The table will have the following transactions :

Transaction #1 , RCPT_SL =1
Transaction #2 , RCPT_SL =2
Transaction #3 , RCPT_SL =3
Transaction #4 , RCPT_SL =5
Transaction #5 , RCPT_SL = 6

## VI. CONCLUSIONS

Autonumber / Auto Sequence Number is a great feature of the modern RDBMS. Commercial RDBMS like Oracle,SQL Server,Db2 etc. and non-commercial like MySql, PostgreSQL etc. all offer this feature. However, it's not advisable to use this feature always while developing an application specially should there be any need for database migration in future. The autonumber module developed and tested by us can be a very easy to use replacement as a database independent autonumber / auto sequence number generator. Though this code is developed and tested using VBA, the same logic is equally applicable for any programming language like Java,C#,C++ etc. which are capable of performing database operations(Eg. read,write etc.).We have tested our module with MS Access, Oracle, MySQL and PostgreySQL. Though we have used the autonumber /auto sequence in many tables , did not face any problem to migrate our database from MSAccess to PostgreSQL since the implementation was not based on database driven default autonumber rather it was implemented by calling the module of autonumber generator as and when required. We also found that numbers generated through our database independent module behaves in the same and/or better way than the database controlled autonumber/auto sequence number.

ACKNOWLEDGMENT

Sincere thanks to the staff and faculty members of Techno India Agartala for actively participating in use and testing of the MIS software developed.